\documentclass[12pt]{article}
\usepackage{amsmath,cite}
\usepackage{amssymb}
\newcommand{\p}[1]{(\ref{#1})}
\newcommand{\be}{\begin{equation}}
\newcommand{\ee}{\end{equation}}
\newcommand{\bea}{\begin{eqnarray}}
\newcommand{\eea}{\end{eqnarray}}
\def\theequation{\arabic{section}.\arabic{equation}}
\begin{document}
\setcounter{page}0
\renewcommand{\thefootnote}{\fnsymbol{footnote}}
\begin{titlepage}
\vskip .7in
\begin{center}
{\Large \bf  Supersymmetric Reducible Higher-Spin Multiplets in Various Dimensions
 } \vskip .7in 
 {\Large 
Dmitri Sorokin$^a$\footnote{e-mail: {\tt  dmitri.sorokin@pd.infn.it }} and
 Mirian Tsulaia$^b$\footnote{e-mail: {\tt  mirian.tsulaia@uwa.edu.au}  }}
 \vskip .4in {$^a$ \it INFN, Sezione di Padova, via F. Marzolo 8, 35131 Padova, Italia} \\
\vskip .2in { $^b$ \it School of Physics $M013$,
The University of
Western Australia, 35 Stirling Highway,
Crawley, Perth, WA 6009, Australia}\\
\vskip .8in
\begin{abstract}

We construct, in $D=3,4,6$ and $10$ space-time dimensions, supersymmetric Lagrangians 
for free massless higher spin fields which belong to reducible representations of the Poincar\'e group.
The fermionic part of these models consists of spinor-tensor fields which are 
totally symmetrical with respect to their tensor indices, while
the bosonic part contains totally symmetric tensor fields as well as the simplest mixed-symmetry fields. A peculiar feature of these models is that
they describe higher- and lower-spin  supermultiplets in different dimensions in a uniform way.

\end{abstract}

\end{center}

\vfill

\end{titlepage}

\renewcommand{\thefootnote}{\arabic{footnote}}
\setcounter{footnote}0

\section{Introduction}
Understanding the relation between string theory and higher spin gauge theory remains a challenging problem. It is based on the hypothesis that
the former is a spontaneously broken phase of the latter, in 
which higher spin fields acquire  masses and string tension is generated. Further connection 
between string theory and higher spin gauge theory is provided by the fact that AdS spaces, which 
are natural backgrounds for interacting higher spin gauge theories, play an important role in 
string theory as the geometric basis for the AdS/CFT correspondence. 
If the connection between 
higher spin gauge theory and string theory is successfully established and understood, it may shed light on the origin and mechanism of tension and mass generation in string theory \footnote{For a review of various aspects and problems of higher spin field theory see e.g. \cite{Vasiliev:1995dn,Vasiliev:1999ba,Vasiliev:2001ur,Bekaert:2003uc,Sorokin:2004ie,Bouatta:2004kk,Bandos:2005rr,Bekaert:2005vh,Francia:2006hp,Fotopoulos:2008ka,Campoleoni:2009je,Francia:2010ap,Taronna:2010qq,Bekaert:2010hw,Sezgin:2012ag,Gaberdiel:2012uj,Giombi:2012ms,Taronna:2012gb,Joung:2012fv,Sagnotti:2013bha,Didenko:2014dwa,Gomez:2014dwa,Rahman:2015pzl,Arias:2016ajh,Sleight:2016hyl,Sleight:2017krf,Leonard:2017uvo,Sorokin:2017irs,Rahman:2017cxk} and Chapter 6.9 of the book \cite{Buchbinder:1995uq}.}.

One of the ways to approach this problem is to take a tensionless (high-energy) limit of string theory and study the structure of its higher-spin excitations which become massless in this limit. This is a highly non-trivial problem on its own.

The simplest case to start with is to perform this limit in flat space-time background and for vanishing string coupling constant, thus restricting the consideration to a free theory.
Acting this way, one obtains consistent Lagrangians and field equations 
for bosonic and fermionic massless fields belonging to various representations of the Poincare group.
These are often called  ``triplets" in the case of totally symmetric tensor fields \cite{Francia:2002pt}  or ``generalized triplets"
\cite{Sagnotti:2003qa} when the system contains tensor fields of mixed symmetry. Various properties of these systems have been studied in 
\cite{Ouvry:1986dv,Bengtsson:1986ys,Bellon:1986ki,Bellon:1986fm,
Henneaux:1988,Pashnev:1989gm,Pashnev:1997rm,Francia:2002pt,Sagnotti:2003qa,Barnich:2005ga,Buchbinder:2006eq,Buchbinder:2007ak,Fotopoulos:2007yq,Alkalaev:2008gi,Fotopoulos:2008ka,Sorokin:2008tf,Fotopoulos:2009iw,Alkalaev:2009vm,Fotopoulos:2010nj,Francia:2010qp,Bekaert:2010hk,Skvortsov:2010nh,Alkalaev:2011zv,Francia:2012rg,Campoleoni:2012th,Asano:2012qn,Asano:2013rka,Bekaert:2015fwa,Buchbinder:2015kca,Agugliaro:2016ngl,Francia:2016weg}.
As a next step, one can promote the resulting system of free fields to the interacting level
\cite{Fotopoulos:2007nm,Sagnotti:2010at,Fotopoulos:2010ay}
and in this way recover
the set of cubic vertices obtained  in 
\cite{Metsaev:1993ap,Metsaev:2005ar,Metsaev:2007rn,Metsaev:2012uy}. 
(See also \cite{Manvelyan:2010jr,Manvelyan:2010je}
and 
\cite{Gross:1987ar,Lindstrom:2003mg,Bonelli:2003kh,Bakas:2004jq,Moeller:2005ez,Fotopoulos:2007nm,Polyakov:2009pk,Bagchi:2016yyf,Lee:2015wwa}
 for work on various aspects of high energy limit in string theory.)

The Lagrangian description of the bosonic higher-spin triplets was generalized to AdS backgrounds in 
\cite{Bengtsson:1990un,Sagnotti:2003qa,Barnich:2005bn,Fotopoulos:2006ci}
and of the fermionic ones in \cite{Sorokin:2008tf,Agugliaro:2016ngl}, though their origin in a tensionless limit of an AdS string is still to be understood.

Another interesting issue, which has not been developed by now and which will be the main goal of this paper, is the construction and study of supersymmetric reducible higher spin systems with the aim of finding their relation to a tensionless limit of superstring theory.

A similarity of higher-spin systems with the spectrum of (super)strings allows one to use techniques and intuition gained in string theory for studying various problems of free and interacting higher spin gauge theories (see e.g. \cite{Sagnotti:2013bha,Rahman:2015pzl} for a review and references). In this paper, these techniques will be applied to the derivation of supersymmetric systems of reducible higher spin multiplets and the construction of their free supersymmetric Lagrangians in flat space-time of dimension $D=3,4,6$ and 10. These values are singled out by the requirement of the on-shell closure of a Poincar\'e supersymmetry algebra, which in what follows  will be assumed minimal (i.e. $N=1$). As we will show the simplest supersymmetric multiplets of the highest spin $s$ consist of fermionic triplets (whose physical fields are contained in symmetric tensor spinors of rank ($s-1$) and generalized bosonic triplets whose physical fields are collected in symmetric tensor fields of rank $s$ and mixed-symmetry tensors characterized by a simple hook Young tableaux $Y(s-1,1)$ of the group $GL(D)$ (with the first row being of rank $s-1$ and the second row of rank 1).
In addition, the systems contain pure gauge and auxiliary fields (not to be confused with those of the off-shell supermultiplets).

To find these supersymmetric higher-spin systems, ~their ~supersymmetry ~transformations ~and Lagrangians, we will use the fact that the fermionic and bosonic triplets (either generilized or not)
can be obtained from the Ramond-Neveu-Schwarz Open String Field Theory by formally taking its   tensionless limit \cite{Sagnotti:2003qa}.
To identify the supersymmetry generator,
one should find an operator which converts the BRST charge of the Ramond (R) sector into the BRST charge
of the Neveu-Schwarz (NS) sector of the theory and vice versa.
In the free Open String Field Theory, this problem was considered  in \cite{Kazama:1986cy} (see also \cite{Terao:1986wg}). It was found that the required operator 
contains an infinite number of string oscillator variables, the situation which naturally appears for a string  of a finite tension.
The supersymmetry algebra closed up to the action of a picture changing operator \cite{Ezawa:1987sk},
which is again characteristic of the NSR formulation \cite{Friedan:1985ge}.

On the other hand, in the formal high energy limit of string theory in which one recovers massless generalized triplets, for their description one uses a 
finite set of oscillator variables whose number may a priori differ in the NS and R sector  (see \cite{Sagnotti:2003qa} for details). Moreover,
the nilpotent BRST charges both in the NS and R sector can be constructed in any space-time dimension $D$.
Therefore it should be possible  to find an analog of the supersymmetry operator given in \cite{Kazama:1986cy}
which contains a finite number of oscillators.
This problem was solved in \cite{Fotopoulos:2008ka} where 
a BRST-invariant Lagrangian and supersymmetry transformations for a system containing a  triplet on the fermionic side (R sector)
and a generalized triplet containing fields with mixed symmetry on the bosonic side (NS sector)
was obtained. However, the structure and properties of the obtained systems were not discussed in \cite{Fotopoulos:2008ka}.

In this paper we elucidate and consider in detail 
the field content, the Lagrangians and supersymmetry transformations 
 of the component fields of the supersymmetric triplet systems introduced in
\cite{Fotopoulos:2008ka}. 
We show how the triplets decompose into various representations of the corresponding  $D$-dimensional
supersymmetry algebra which closes on-shell (for $D=3,4,6$ and 10) and is not plagued by the problem of picture changing.

Since our formalism closely follows the one of superstring theory, they have some features in common. For instance,
the lowest spin supermultiplet in Type I $D=10$ open string theory is the Maxwell multiplet. It is well known that, although the free Lagrangian for
an Abelian vector and a spinor field is invariant under $N=1$ supersymmetry transformations in any space-time dimension $D$,
the closure of the supersymmetry algebra requires that $D=3,4,6$ and $10$. 
We will see that a similar restriction persists for our  models
\footnote{Since we consider massless fields the case of $D=3$ appears somewhat trivial (see \cite{Buchbinder:2017rto} for a review of the construction
of supersymmetric massive higher spin fields in $D=3$ flat and $AdS_3$ spaces). Therefore, although our consideration is also valid for $D=3$,
we will only discuss $D=4,6$ and $10$.}.

Let us note that  in $D=4$ the problem of supersymmetric formulation of reducible higher spin systems was considered earlier in
\cite{Bellon:1986ki,Bellon:1986fm}. The fermionic sector of their models consisting of the totally symmetrical spinor-tensors is the same as in our case but in the bosonic sectors, the corresponding Lagrangians and supersymmetry transformations are different. In short, the systems of \cite{Bellon:1986ki,Bellon:1986fm}
are  a higher-spin generalization of the chiral  $D=4$ supermultiplet, while we will deal with a higher--spin generalization of the $D=4$ vector supermultiplet.

The paper is organized as follows.
In Section \ref{General} we briefly review 
gauge invariant systems describing bosonic and fermionic fields,
which belong to reducible representations of the Poincar\'e group with mixed symmetries
(generalized triplets) \cite{Sagnotti:2003qa}. We present corresponding Lagrangians, gauge transformations, field equations
and BRST charges, valid in any number of space-time dimensions and for any finite number of oscillator variables.

In  Section \ref{Fieldcontent} we consider the simplest sets of reducible higher spin systems, which form 
minimal $N=1$ supermultiplets in $D=3,4,6$ and 10. They consist of a fermionic triplets (spinor fields which are symmetric tensors of the Lorentz group) and generalized bosonic triplets containing symmetric tensors and fields of the simplest mixed symmetry of the $Y(s-1,1)$ type.
We present the BRST-invariant Lagrangians, field equations, and gauge transformations of these systems.

In particular, in Section \ref{Supersymmetry} we review the form of an operator which transforms the BRST charge
of the fermionic sector into the BRST charge of the bosonic sector and thus is supposed to generate their  supersymmetry transformations \cite{Fotopoulos:2008ka}.

Section \ref{Supermultilets} contains our main results.
We start by discussing how the $N=1$ Maxwell supermultiplet is obtained within this formalism. This serves as a useful warming up example and a preparation for the consideration of supergravity and higher-spin supermultiplets.

The next example deals with linearized supergravity multiplets. It already has interesting peculiarities.
We will see that while in $D=10$ after integrating out auxiliary fields we recover the irreducible $N=1$, $D=10$ (linearized) supergravity multiplet, in the lower dimensional cases ( $D=4$ and 6) the systems become reducible and split into linearized supergravity and tensor multiplets {}\footnote{One comment is in order here. As we mentioned,
the systems under consideration have been deduced by taking a tensionless limit of open superstring field theory. The appearance of the supergravity multiplet
is due to the fact that in this limit the spin-$2$ field becomes massless and is described by the Fierz-Pauli Lagrangian. It can thus be naturally identified with a graviton at least at the free level.}. 

Finally, we present a general case of simple reducible higher-spin supermultiplets.

In conclusion we discuss open problems and possible developments of our results.

\section{General set up} \setcounter{equation}0 \label{General}

Let us briefly recall the description of higher spin fields transforming under reducible representations of the
Poincar\'e group in the BRST approach \cite{Francia:2002pt,Sagnotti:2003qa,Campoleoni:2012th}, see also \cite{Fotopoulos:2008ka} for a review. 

The BRST formalism used below is very similar to the RNS formulation of the open superstring.
In particular, one introduces bosonic and fermionic oscillator operators as well as bosonic and fermionic ghost variables to create states of different spins.

Bosonic oscillators $\alpha^\mu_k$,  fermionic ghosts $c_k$  and antighosts $b_k$ obey the following (anti)commutation relations
\begin{equation}\label{B4}
[\alpha^\mu_k, \alpha^{\nu +}_{l} ] = \delta_{kl} \eta^{\mu \nu},
\quad \{ c^{+}_k, b_l \} = \{ c_k, b^{+}_l \} = 
 \delta_{kl}\,,
\end{equation}
$$
\{ c_0 , b_0 \}=1,
$$
where $k,l=1,...,\infty$ are positive integer numbers. 
The ghost number of $c^{+}_k$ is $+1$, the ghost number of  $b^{+}_k$ is $-1$
and the ghost number of  $\alpha_{k}^{\mu +}$  is $0$.

The fermionic oscillators $\psi^\mu_r$ and the bosonic ghosts $\gamma_r$
and antighosts ${\beta}_r$ obey the following  anti-commutation relations
\begin{equation}
\{ \psi^\mu_r, \psi^{\nu +}_s \} \ = \ \delta_{rs}\, \eta^{\mu \nu} \ ,
\qquad
[\gamma_r , {\beta}_{s}^+] \ = [\gamma^{+}_r , {\beta}_{s}]  = \ i \, \delta_{rs}\,.
\end{equation}
The ghost numbers of $\gamma^{+}_r$, $\beta^{+}_r$ and  $\psi^{\mu+}_r$ are, respectively, $+1$,  $-1$ and $0$.
All ``daggered"oscillators are Hermitian conjugated to the corresponding ``undaggered" ones. 
The oscillators with the subscript "$0$" are Hermitian self-conjugate.

Keeping the terminology adopted in string theory, we call the NS sector the set of string states in which the indices of the fermionic oscillators and the corresponding
bosonic (anti)ghosts assume half integer values, namely $r,s= \frac{1}{2},..., M + \frac{1}{2}$.
The NS sector consists of space-time bosonic fields.
In a similar way, we call the R sector the set of string states in which the indices of the fermionic oscillators and the corresponding bosonic (anti)ghosts 
assume non-negative integer values $r,s= 0,1,..., \infty$. The R sector consists of space-time fermion fields.

Let us note that the number of oscillators $K,M$ and $N$ can be a priori arbitrary and independent from each other.
Fields are expanded in terms of creation operators in both sectors.
The Clifford vacua are defined as follows. In the NS sector the vacuum state is 
defined as follows
\be\label{vacuumns-1}
\alpha^\mu_k|0_{NS} \rangle = \psi^\mu_k|0_{NS}\rangle = 0, 
\ee
\be\label{vacuumns-2}
c_k|0_{NS} \rangle = b_k|0_{NS} \rangle = \beta_k|0_{NS}\rangle = \gamma_k|0_{NS} \rangle =b_0 |0_{NS} \rangle=0.
\ee
The vacuum in the R sector is defined in a similar way
\be\label{vacuumr-1}
\alpha^\mu_k|0_R \rangle = \psi^\mu_k|0_R \rangle = 0, \quad k>0
\ee
\be\label{vacuumr-2}
c_k|0_R \rangle = b_k|0_R \rangle = \beta_k|0_R \rangle = \gamma_k|0_R \rangle =b_0 |0_R \rangle =0, \quad k>0
\ee
\be \label{vacuumr-3}
\beta_0 |0_R \rangle =0.
\ee
As in superstring theory, $\psi^\mu_0$ are usually associated with gamma matrices, therefore the Clifford vacuum state in the R-sector has a
spinorial index (i.e., we have $| 0_R \rangle_a$ and ${}^a\langle 0_R|$), which we will not write explicitly.

A general field either in  the $NS$ or $R$ sector, which we will call higher-spin functional, 
is expanded in terms of the creation operators introduced above and  the components of this expansion
are higher spin fields (physical and auxiliary).

Let us now introduce differential operators.
In the $\alpha$-oscillator parts of the both sectors we will deal with  
\be\label{l}
l_0 = p\cdot p, \qquad l_k = p \cdot \alpha_k, \qquad l^{+}_k = p \cdot \alpha^{+}_k,
\ee
where $p_\mu = -i \partial_\mu$, `dot' stands for the trace (i.e. $A \cdot B= A^\mu B_\mu$) and $\partial  A$ will denote a symmetrized derivative. For example if $A$  is a vector, then $\partial  A \equiv \partial_\mu A_\nu + \partial_\nu A_\mu$. 

In addition to the above operators, in the NS sector we have 
\be \label{1111}
g_r = p \cdot \psi_r, \qquad g^{+}_r = p \cdot \psi^{+}_r, \qquad r+\frac{1}{2} \in \mathbb{N} , \quad
\mathbb{N}=1,2,...
\ee 
and in the R sector
\be \label{2222}
g_0 = p \cdot \psi_0, \quad g_r = p \cdot \psi_r, \quad g^{+}_r = p \cdot \psi^{+}_r\,, \qquad r \in \mathbb{N}\, . \quad
\ee 
Obviously, $g_0$ is the Dirac operator and $l_0$ is the d'Alembertian.

The action of the operators $l_k$  (defined in \eqref{l}) on the states created by $\alpha^{\mu+}_k$  results in taking the divergence of the corresponding tensorial fields.
Namely, since the vacuum is defined as in \p{vacuumns-1} and \p{vacuumr-1}, the expansion of any state
in series of, say, oscillator $\alpha^{\mu +}_1$ (ignoring the other creation operators for a moment) has the form
\be \label{bbbb}
| \Phi \rangle = \frac{1}{m!}  \Phi_{\mu_1 \mu_2,..., \mu_m}(x)\alpha^{\mu_1 +}_1 \alpha^{\mu_2 +}_1
...\alpha^{\mu_m +}_1 
 | 0_{NS/R} \rangle.
\ee
Then  acting on \p{bbbb} with the operator $l_1$ and performing the normal ordering of
$\alpha^{\mu}_1$ and $\alpha^{\mu +}_1$ one can see that the resulting state is the divergence of the initial one
\be\label{dbbbb}
| \Phi \rangle = \frac{1}{(m-1)!}  \partial^{\mu_1}\Phi_{\mu_1 \mu_2,..., \mu_m}(x) \alpha^{\mu_2 +}_1
...\alpha^{\mu_m +}_1 
 | 0_{NS/R} \rangle.
\ee
Similarly,
$g_r$ acts as the divergence on the states created by
 $\psi^{\mu +}_r$.
The operators $l^{+}_i$ produce symmetrized derivatives of the tensor fields associated with the states created by $\alpha^{\mu +}_i$ and finally 
 $g^{+}_r$ act as the antisymmetrized derivatives of the tensor fields associated with the states created by the fermionic oscillators $\psi^{\mu +}_r$.

In the NS sector the BRST charge is
 \be \label{BRSTNS} Q_{NS} \ = \ c_0 \,
l_0 \ + \ \tilde Q_{NS} \ - \ M_{NS} \, b_0 \ , \ee with
\begin{eqnarray}
&& \label{TQ} \tilde Q_{NS}\ = \ \sum_{ k,r } \; \left [\;
c_{k}^+ \, l_k   + c_{k} \, l_k^+ +  \gamma_{r}^+\, g_r +  \gamma_{r}\, g_r^+ \right ]\ , \nonumber \\
\label{TN} && 
M_{NS}\ = \  \sum_{k,r }
\; \left [ \,  \, c_{k}^+\, c_k  + \gamma_{r}^+ \, \gamma_r \,
\right ] \ ,
\end{eqnarray}
and in the R sector
 \be \label{BRSTR}
Q_R \ = \ c_0 \, l_0  +  \gamma_0 \, g_0  +  \tilde Q_R  -  M_R
b_0  -  \frac{1}{2}\, \gamma_0^2 \, b_0 \ , \ee where $\tilde
Q_R$ and $M_R$ are defined similarly to (\ref{TN}) but with the difference that now the sums are taken over the
 integer modes of the fermionic oscillators and
the bosonic (anti)ghosts. The BRST charges are identically
nilpotent
\be
Q^2_{NS}= Q^2_{R}=0
\ee
independently of the space-time dimension $D$ and of the numbers of the oscillators $K,M$ and $N$.

Let us note that, since the states are series expansions in both the fermionic and bosonic creation operators,
the proper definition of parity of the higher-spin functionals is needed.
This is determined by the GSO projector which in  the NS sector is
 \be \label{GSONS}
 P_{NS} \ = \
\frac{1}{2} \, \left[ 1 \ - \ (-1)^{\sum(\psi_p^\dagger \; \psi_p \,
+\, i \gamma_p^\dagger \; \beta_p \, - \, i\; \gamma_p\;
\beta_p^\dagger)}\right] \ee
and in the R sector has the following form 
 \be \label{GSOR}
 P_R \ = \ \frac{1}{2}\,\left[ 1\ +\
\gamma_{*}\, (-1)^{\sum(\psi^\dagger_r \; \psi_r \, + \, i\;
\gamma^\dagger_r \; \beta_r \, - \, i \; \gamma_r \;
\beta^\dagger_r \, + \, i \; \gamma_0 \; \beta_0)}\right] \ , \ee
where $\gamma_{*}$ is a chirality matrix.

The quadratic NS-sector Lagrangian is
\be \label{falphal}
L_{NS}= \int d c_0 \langle \Phi_{NS} | Q_{NS} | \Phi_{NS} \rangle.
\ee
It is invariant under the following gauge transformations
\be \label{GAUGENS}
\delta |\Phi_{NS} \rangle = Q_{NS} | \Lambda_{NS} \rangle
\ee
due to the nilpotency of the BRST charge.
Writing the NS field and the gauge parameter as polinomials of the
fermionic ghost zero-mode 
\begin{equation} \label{HSFNS}
 |\Phi^{NS} \rangle \ = \ |\Phi_1^{NS}\rangle \ + \ c_0|\Phi_2^{NS}
\rangle \ ,  
\end{equation}
\begin{equation}
 |\Lambda^{NS} \rangle \ = \ |\Lambda_1^{NS} \rangle
\ + \ c_0 |\Lambda_2^{NS} \rangle  \ , 
\end{equation}
and making use of the form of the BRST charge
(\ref{BRSTNS}), one rewrites the Lagrangian \p{falphal} as follows
\be \label{LNS}
L_{NS}=\langle \Phi^{NS}_1|l_0|\Phi^{NS}_1\rangle -
 \langle \Phi^{NS}_2|\tilde Q_{NS}|\Phi^{NS}_1\rangle -
 \langle \Phi^{NS}_1|\tilde Q_{NS}|\Phi^{NS}_2\rangle +
 \langle \Phi^{NS}_2|M_{NS}|\Phi^{NS}_2\rangle\,. 
\ee
Varying this Lagrangian one gets the following equations of motion
\begin{eqnarray} \label{EMNS1}
&& l_0 |\Phi_1^{NS} \rangle \ - \ \tilde Q_{NS}|\Phi_2^{NS} \rangle \ = \ 0
\ , \nonumber  \\
&& \tilde Q_{NS} |\Phi_1^{NS} \rangle \ - \ M_{NS} |\Phi_2^{NS}
\rangle \ =\ 0 \ . \label{EMNS2} \end{eqnarray}
The
gauge transformations \p{GAUGENS} of the fields $|\Phi_1^{NS} \rangle $ and $|\Phi_2^{NS} \rangle $ are
\begin{eqnarray}
&& \delta |\Phi_1^{NS} \rangle \ = \ \tilde Q_{NS}
|\Lambda_1^{NS} \rangle \ - \ M_{NS} |\Lambda_2^{NS} \rangle \ , \nonumber \\
\label{NSG3}
&& \delta |\Phi_2^{NS} \rangle \ =\
l_0 |\Lambda_1^{NS} \rangle \ - \ \tilde Q_{NS}|\Lambda_2^{NS} \rangle \ .
\end{eqnarray}
The R sector is more complicated, due to the presence of the
bosonic ghost zero mode $\gamma_0$. In general, $|\Phi^R \rangle \ $ is an infinite series in powers of $\gamma_0$.  However, one can consider a
truncated field
\begin{equation} \label{truncated}
|\Phi^R \rangle \ = \ |\Phi_1^R\rangle \ + \ \gamma_0 \,  |\Phi_2^R \rangle
\ + \ 2 \, c_0 \, g_0\, | \Phi^R_2\rangle \ ,
\end{equation}
without affecting the physical spectrum \cite{Kazama:1986cy}. The truncated sector still possesses a relevant portion of gauge symmetry\footnote{In other words one can prove that
using the BRST gauge symmetry $\delta | \Phi_R \rangle = Q_R | \Lambda_R \rangle$ of the equation
$Q_R | \Phi_R \rangle =0$ one can  gauge fix the functional $ | \Phi_R \rangle$ to the form \p{truncated}.}.
The resulting consistently truncated field equations
\begin{eqnarray} \label{EMR1}
&& g_0 \, |\Phi_1^R \rangle \ + \ \tilde Q_{R}|\Phi_2^R \rangle \ =\ 0 \ ,
\nonumber \\
 && \tilde Q_{R} \, |\Phi_1^R \rangle \ - \ 2 \, M_R
\, g_0 \, |\Phi_2^R \rangle \ = \ 0 \ ,
\end{eqnarray}
are invariant under the following gauge transformations
\begin{eqnarray} \label{GTR1}
&& \delta\, |\Phi_1^R\rangle \ = \ \tilde Q_R |\Lambda_1^R \rangle
\ +\ 2 \,M_R \, g_0 \, |\Lambda_2^R \rangle \ , \label{GT1}
\nonumber \\
\label{GTR2}
&& \delta|\Phi_2^R\rangle \ = \ g_0 \, |\Lambda_1^R \rangle \ - \ \tilde
Q_R \, |\Lambda_2^R \rangle \ . \label{GT2}
\end{eqnarray}
The field equations \p{EMR1} are obtained from the Lagrangian
\be
L_R= \langle \Phi^{R}_1|g_0|\Phi^{R}_1\rangle +
 \langle \Phi^{R}_2|\tilde Q_{R}|\Phi^{R}_1\rangle +
 \langle \Phi^{R}_1|\tilde Q_{R}|\Phi^{R}_2\rangle -
2 \langle \Phi^{R}_2|M_R g_0|\Phi^{R}_2\rangle\,.
\ee

A way to establish   supersymmetry between the fields of the NS and R sector is to find
an operator $\mathcal Q$ which transforms the BRST charges $Q_R$ and $Q_{NS}$ into each other 
\be \label{DW}
Q_R\, \mathcal Q = \mathcal Q\, Q_{NS}.
\ee
Such an operator has been considered in the context of open string field theory  in
\cite{Kazama:1986cy}. 
The condition \eqref{DW} is the consequence of the requirement that the total Lagrangian 
\be \label{TSCH}
L_{tot.}= \langle \Phi_{NS} |Q_{NS}| \Phi_{NS} \rangle + \langle \Phi_R |Q_R| \Phi_R \rangle
\ee
is  invariant under  
\be \label{LL}
\delta \langle \Phi_{NS}|  = \langle \Phi_R| \epsilon \mathcal Q , \qquad \delta | \Phi_{R} \rangle = \epsilon \mathcal Q|\Phi_{NS} \rangle.
\ee 
where $\epsilon$ is a Grassmann-odd constant parameter of the supersymmetry transformations. 

\section{Field content of the simplest $N=1$ supersymmetric reducible higher-spin systems} \setcounter{equation}0 \label{Fieldcontent}

Let us consider the simplest case in which the fields in the $R$ sector do not depend  on the
fermionic oscillators $\psi^{\mu +}_r$. These are the
fermionic higher-spin triplets  considered in flat space-time \cite{Bellon:1986ki}, \cite{Francia:2002pt,Sagnotti:2003qa} and in $AdS_D$  \cite{Sorokin:2008tf,Agugliaro:2016ngl}.

The higher-spin functional of the R sector \p{truncated} associated with the fermionic triplet contains only the oscillators $\alpha_1^{\mu +}$, $c_1^+$ and $b_1^+$, namely
\begin{eqnarray}\label{Rtrip}
&&|\Phi_1^R\rangle  = \ \frac{1}{n!}\ \Psi_{\mu_1\mu_2\, ...\,
\mu_n}(x)\, \alpha^{\mu_1+}_1 \, \alpha^{\mu_2 +}_1 \, ...\,
\alpha^{\mu_n +}_1\, |0_R \rangle \nonumber \\ \nonumber && \qquad
\qquad +\ \frac{ c_{1}^+ b_{1}^+}{(n-2)!}\ \Sigma_{\mu_1\mu_2\, ...\,
\mu_{n-2}}(x)\, \alpha^{\mu_1+}_1 \, \alpha^{ \mu_2+}_1 \, ..\,
\alpha^{ \mu_{n-2}+}_1 \, |0_R \rangle \ ,
\\
&&|\Phi_2^R\rangle \ = \  \frac{b_{1}^+}{\sqrt{2} (n-1)!}\
\chi_{\mu_1\mu_2\, ...\, \mu_{n-1}}(x) \, \alpha^{\mu_1+}_1 \,
\alpha^{ \mu_2+}_1 \, ...\, \alpha^{\mu_{n-1}+}_1\, |0_R \rangle
\end{eqnarray}
or in a more compact form
\bea \label{Rtrip-c}
&&|\Phi_1^R\rangle = |\Psi \rangle + c_{1}^+ b_{1}^+ | \Sigma \rangle, \\ 
&& |\Phi_2^R\rangle=  b_1^+ | \chi \rangle.
\eea
The monomials of the oscillators $\alpha^{\mu}$ define the
spinor-tensor fields $\psi$, $\chi$ and $\Sigma$ of rank $(n+1/2)$, $(n-1/2)$
and $(n-3/2)$, respectively, which are totally
symmetric in their tensor indices. Substituting these expressions into the
field equations (\ref{EMR1}) one gets the fermionic triplet equations of motion
\be \label{Er-1}
g_0 | \Psi \rangle  + l^+ | \chi \rangle=0\,,
\ee
\be \label{Er-2}
l  | \Psi \rangle -  l^+ | \Sigma \rangle + 2 g_0 | \chi \rangle =0\,,
\ee
\be \label{Er-3}
g_0 | \Sigma \rangle + l  | \chi \rangle=0\,,
\ee
with the corresponding Lagrangian having the following form
\bea \label{RR-1}
L_R &=&\langle \Psi |g_0 | \Psi \rangle -  \langle \Sigma |g_0 | \Sigma \rangle +  
\langle \Psi |l^+ | \chi \rangle
- \langle \Sigma |l  | \chi \rangle \\ \nonumber
&&+  \langle \chi |l  | \Psi \rangle -  \langle \chi |l^+ | \Sigma \rangle +  2 \langle \chi |g_0 | \chi \rangle\,.
\eea
The BRST gauge invariance of this Lagrangian involves an unconstrained parameter
\begin{equation}
|\Lambda^R_1 \rangle \ =\ \frac{ib_{1}^+}{(n-1)!}\
\tilde \lambda_{\mu_1\mu_2...\mu_{n-1}}(x) \, \alpha^{+\mu_1}_1 \,
\alpha^{+\mu_2}_1 ... \alpha^{+\mu_{n-1}}_1\, |0\rangle \ ,
\end{equation}
or simply
\be
|\Lambda^R_1 \rangle = b_1^+ | \tilde \lambda \rangle\,,
\ee
and the gauge transformations are
\begin{eqnarray}
&& \delta |\Psi \rangle \ = \  l^+ \, |\Lambda^R_1 \rangle \ , \nonumber \\
&& \delta |\Sigma \rangle \ = \  l \, |\Lambda^R_1 \rangle \ ,
\nonumber \\
&& \delta |\chi \rangle \ = \ -g_0 \, |\Lambda^R_1 \rangle \ .
\end{eqnarray}
In order to pass to the conventional field-theoretical description of this fermionic system  we identify $\psi^\mu_0$ with the gamma-matrices as follows 
\be
\psi^\mu_0=\frac 1{\sqrt{2} }\gamma^\mu  
\ee
and use the equations above to rewrite the Lagrangian in the following form (in which the contraction of the tensorial indices is assumed)
\bea \label{RR-2}
L_R &=& -i \bar \Psi \gamma^\mu \partial_\mu \Psi -i  n  \bar \Psi \partial \chi + i n
 \bar \chi \partial \cdot \Psi +
i n (n-1)\bar \Sigma \gamma^\mu \partial_\mu \Sigma \\ \nonumber
&+& i n \bar \chi \gamma^\mu \partial_\mu \chi - i n (n-1)\bar \chi \partial \Sigma + i n (n-1)\bar \Sigma  \partial \cdot \chi\,.
\eea
The Lagrangian is invariant under the gauge transformations
\begin{eqnarray} \label{gtr-f-1}
&& \delta \Psi \ = \ \partial \, \tilde \lambda \ , \nonumber \\
&& \delta \Sigma \ = \ \partial \cdot \tilde \lambda \ ,
\nonumber \\
&& \delta  \chi \ = \ - \gamma^\nu \partial_\nu \tilde  \lambda \ .
\end{eqnarray}
From the Lagrangian \p{RR-2} one obtains the equations of motion
\begin{eqnarray}
&& \gamma^\nu \partial_\nu \Psi \ + \ \partial \chi =0 \ , \nonumber \\
&& \partial \cdot \Psi \ - \ \partial \Sigma
+ \gamma^\nu \partial_\nu \chi =0 \ , \nonumber \\
&& \gamma^\nu \partial_\nu \Sigma \ +\
\partial \cdot \chi =0. \label{fermitriplet}
\end{eqnarray}

For the fields in the NS sector we require that they are generated by the oscillators
$\alpha_1^{\mu +}, c_1^+, b_1^+$ as well as by  $\psi_\frac{1}{2}^{\mu +}, \beta_{\frac{1}{2}}^{+}$ and  $\gamma_{\frac{1}{2}}^{+}$. The reason for this choice is that the number of the on-shell physical degrees of freedom of this bosonic system matches that of the fermionic triplet, as we show in detail in Appendix B. The equations of motion in the NS sector \p{EMNS1} fix the following form of the higher-spin functional
\begin{eqnarray}\label{NStrip}
&& |\Phi_1^{NS} \rangle \ = \ \frac{1}{n!}\ \phi_{ \nu, \mu_1\mu_2\, ...\,
\mu_n }(x)\, \psi^{+\nu}_{\frac{1}{2}}  \alpha^{\mu_1+}_1 \, \alpha^{\mu_2 +}_1 \, ...\,
\alpha^{\mu_n +}_1   \, |0_{NS} \rangle \nonumber \\ \nonumber && \qquad
\qquad +\ \frac{ c_{1}^+ b_{1}^+}{(n-2)!}\ D_{ \nu,\, \mu_1\mu_2\, ...\,
\mu_{n-2}  }(x)\, \psi^{\nu+}_{\frac{1}{2}} \alpha^{\mu_1+}_1 \, \alpha^{ \mu_2+}_1 \, ..\,
\alpha^{ \mu_{n-2}+}_1  \, |0_{NS} \rangle \\ \nonumber && \qquad
\qquad +\ \frac{ \gamma_{1/2}^+ b_{1}^+}{(n-1)!}\ B_{\mu_1\mu_2\, ...\,
 \mu_{n-1}}(x)\, \alpha^{\mu_1+}_1 \, \alpha^{ \mu_2+}_1 \, ..\,
\alpha^{ \mu_{n-1}+}_1  \, |0_{NS} \rangle \\ \nonumber && \qquad
\qquad +\ \frac{ i c_{1}^+ \beta_{1/2}^+}{(n-1)!}\ A_{\mu_1\mu_2\, ...\,
 \mu_{n-1}}(x)\, \alpha^{\mu_1+}_1 \, \alpha^{ \mu_2+}_1 \, ..\,
\alpha^{ \mu_{n-1}+}_1  \, |0_{NS} \rangle \,,
\\
&&|\Phi_2^{NS}\rangle \ = \ \ \frac{ib_{1}^+}{\; (n-1)!}\
C_{ \nu, \mu_1\mu_2\, ...\, \mu_{n-1}  }(x) \, \psi^{\nu+}_{\frac{1}{2}}\, \alpha^{\mu_1+}_1 \,
\alpha^{ \mu_2+}_1 \, ...\, \alpha^{\mu_{n-1}+}_1 \, |0_{NS} \rangle \\ \nonumber && \qquad
\qquad  \  +\ \frac{\beta_{1/2}^+}{\; n!}\
E_{\mu_1\mu_2\, ...\, \mu_{n}}(x) \, \alpha^{\mu_1+}_1 \,
\alpha^{ \mu_2+}_1 \, ...\, \alpha^{\mu_{n}+}_1\, |0_{NS} \rangle  \\ \nonumber && \qquad
\qquad  \  +\ \frac{c_1^+ b_1^+\beta_{1/2}^+}{\; (n-2)!}\
F_{\mu_1\mu_2\, ...\, \mu_{n-2}}(x) \, \alpha^{\mu_1+}_1 \,
\alpha^{ \mu_2+}_1 \, ...\, \alpha^{\mu_{n-2}+}_1\, |0_{NS} \rangle \,.
\end{eqnarray}
Note that in \eqref{NStrip} there is no any symmetry between the indices $\nu$ and $\mu$. This is indicated by a comma separating the indices.
The expressions \eqref{NStrip} can be written in a more compact form as follows
\bea
&&|\Phi_1^{NS}\rangle= |\phi \rangle + c_{1}^+ b_{1}^+ | D \rangle + \gamma_{1/2}^+ b_{1}^+ | B \rangle + 
c_{1}^+ \beta_{1/2}^+ | A \rangle, \\ \nonumber
&&|\Phi_2^{NS}\rangle= b_{1}^+ | C \rangle + \beta_{1/2}^+ | E \rangle + c_1^+ b_1^+\beta_{1/2}^+ | F \rangle.
\eea
We will also use the following notation. A field $\Phi_{k,n}$ has $n$ symmetries indices $\mu$ and  $k=0$ or 1  index $\nu$. There is no symmetry between $\mu$ and
$\nu$.
In this notation the reducible system of higher-spin fields of the $NS$ sector 
consists of the mixed symmetry  fields $\phi_{1,n}$ and
$D_{1,n-2}$, the symmetric fields $A_{0, n-1}$ and $B_{0,n-1}$, and  the auxiliary fields 
 $C_{1,n-1}$, $E_{0,n}$ and $F_{0,n-2}$ which can be excluded by solving their equations of motion.

The parameters of the gauge transformations \p{NSG3} have the following form
\begin{eqnarray}\label{NStrip1}
&& |\Lambda_1^{NS} \rangle \ = \ \frac{i b_1^+}{(n-1)!}\ \lambda_{\nu, \mu_1\mu_2\, ...\,
\mu_{n-1} }(x)\, \psi^{\nu+}_{\frac{1}{2}}\, \alpha^{\mu_1+}_1 \, \alpha^{\mu_2 +}_1 \, ...\,
\alpha^{\mu_{n-1} +}_1   \, |0_{NS} \rangle \nonumber \\ \nonumber && \qquad
\qquad +\ \frac{ \beta_{1/2}^+}{n!}\ \rho^{(1)}_{\mu_1\mu_2\, ...\,
\mu_n}(x)\, \alpha^{\mu_1+}_1 \, \alpha^{ \mu_2+}_1 \, ..\,
\alpha^{ \mu_n +}_1 \, |0_{NS} \rangle \\ \nonumber && \qquad
\qquad +\ \frac{ c_{1}^+b_1^+ \beta_{1/2}^+}{(n-2)!}\rho^{(2)}_{\mu_1\mu_2\, ...\,
\mu_{n-2}}(x)\, \alpha^{\mu_1+}_1 \, \alpha^{ \mu_2+}_1 \, ..\,
\alpha^{ \mu_{n-2} +}_1 \, |0_{NS} \rangle \,,
\\
&&|\Lambda_2^{NS}\rangle \ = \ \ \frac{ib_{1}^+  \beta_{1/2}^+}{\; (n-1)!}\
\tau_{\mu_1\mu_2\, ...\, \mu_{n-1} }(x) \, \alpha^{\mu_1+}_1 \,
\alpha^{ \mu_2+}_1 \, ...\, \alpha^{\mu_{n-1}+}_1  \, |0_{NS} \rangle 
\end{eqnarray}
or
\bea
&&|\Lambda_1^{NS} \rangle = b_1^+ |\lambda \rangle + \beta_{1/2}^+ |\rho^{(1)} \rangle + \beta_{1/2}^+ c_1^+ b_1^+ | \rho^{(2)} \rangle , \\ \nonumber
&&|\Lambda_2^{NS} \rangle =  b_1^+\beta_{1/2}^+ |\tau \rangle .
\eea
For the component fields we  have
\begin{eqnarray} \label{NSNSGT} \nonumber
\delta | \phi \rangle &=& l^+ | \lambda \rangle  + i g^+ | \rho^{(1)} \rangle ,\\ \nonumber
\delta | D \rangle &=& l | \lambda \rangle  + i g^+ | \rho^{(2)}\rangle , \\  \nonumber
\delta | A \rangle &=&- l^+ | \rho^{(2)} \rangle  + l | \rho^{(1)} \rangle - | \tau \rangle ,\\  \nonumber
\delta | B\rangle &=&- g| \lambda \rangle  -i | \tau \rangle ,\\
\delta | C \rangle &=& l_0| \lambda  \rangle  + i g^+ | \tau \rangle ,\\  \nonumber
\delta | E \rangle &=& l_0| \rho^{(1)} \rangle  - l^+ | \tau \rangle , \\  \nonumber
\delta | F \rangle &=& l_0| \rho^{(2)} \rangle  -  l | \tau \rangle .
\end{eqnarray}
Finally the Lagrangian \p{LNS} takes the following form for the component fields of the reducible higher-spin system
\begin{eqnarray} \label{NSNS}
L_{NS}&&= \langle \phi | l_0 | \phi \rangle -  \langle D |l_0 | D\rangle + i  \langle B| l_0 | A\rangle - i  \langle A | l_0 | B\rangle \\ \nonumber
&& + \langle C| l^+ | D\rangle + i  \langle E| l^+ | B \rangle -  \langle C| l | \phi \rangle 
- i \langle F | l| B \rangle \\ \nonumber
&&+ i  \langle C | g^+|A \rangle + i  \langle E| g | \phi \rangle-i  \langle F | g| D \rangle \\ \nonumber
&& + \langle D | l| C\rangle - i  \langle B | l | E \rangle - \langle \phi | l^+| C\rangle + i  \langle B | l^+ | F \rangle 
- i  \langle A| g| C \rangle \\ \nonumber
&&- i  \langle \phi| g^+| E \rangle + i  \langle D | g^+ | F \rangle \\ \nonumber
&&+ \langle C| C \rangle + \langle E|  E \rangle - \langle F|  F\rangle .
 \end{eqnarray}
 The equations of motion which follow from this Lagrangian are
\begin{eqnarray}  \nonumber
&& l_0 | \phi \rangle - l^+ | C \rangle - i g^+ | E \rangle =0 ,\label{Ens-1} \\  \nonumber
&&  l_0 | D\rangle  -l | C \rangle - i g^+ | F\rangle =0, \label{Ens-2} \\  \nonumber
&& l_0 | A \rangle - l | E \rangle + l^+ | F \rangle =0, \label{Ens-3} \\  \nonumber
&& l_0 | B\rangle +g | C \rangle =0 ,\label{Ens-4} \\
&& | C \rangle + l^+ | D \rangle - l | \phi \rangle + i g^+ | A \rangle =0, \label{Ens-5} \\  \nonumber
&& | E\rangle +i l^+ | B \rangle +i g| \phi \rangle =0, \label{Ens-6} \\  \nonumber
&& | F \rangle +  il | B \rangle + ig | D \rangle =0. \label{Ens-7}
\end{eqnarray}
As one can see, the fields $| C \rangle$, 
$| E \rangle $ and $| F \rangle $ are expressed in terms of the other fields. These expressions can be susbstituted back into the Lagrangian, which will then describe a system of four fields $| \phi \rangle$, $| D \rangle$, $| A \rangle $ and $| B \rangle $.

 Let us present the Lagrangian \p{NSNS} in a slightly different form, which is
 convenient for checking its supersymmetry invariance. In particular, upon `integrating out' the oscillator 
 $\psi^{\nu +}$ one gets
\begin{eqnarray} \label{tn-2}
L_{NS}&&= \langle \phi^\nu | l_0 | \phi_\nu \rangle - \langle D^\nu |l_0 | D_\nu\rangle 
+ i  \langle B| l_0 | A\rangle - i  \langle A | l_0 | B\rangle \\ \nonumber
&& + \langle C^\nu| l^+ | D_\nu \rangle + i  \langle E| l^+ | B \rangle -  \langle C^\nu| l | \phi_\nu \rangle 
- i \langle F | l| B \rangle +   \langle C^\nu| \partial_\nu|A \rangle \\ \nonumber
&&+ \langle E| \partial^\nu | \phi_\nu \rangle
-  \langle F | \partial^\nu | D_\nu \rangle \\ \nonumber
&& + \langle D^\nu | l| C_\nu \rangle - i  \langle B | l | E \rangle - \langle \phi^\nu | l^+| C_\nu\rangle +
 i  \langle B | l^+ | F \rangle  
-  \langle A^\nu| \partial_\nu | C \rangle \\ \nonumber
&&-  \langle \phi^\nu| \partial_\nu| E \rangle +  \langle D^\nu | \partial_\nu | F \rangle \\ \nonumber
&&+ \langle C^\nu| C_\nu \rangle + \langle E|E \rangle - \langle F| F\rangle, 
 \end{eqnarray}
or in the (compact) tensor notation
\bea \label{tn}
L_{NS}&&=  - \phi^\nu \Box \phi_\nu + n(n-1)D \Box D + nB\Box A  + nA \Box  B \\ \nonumber
&& -2n B \partial \cdot E + 2n (n-1)D^\nu \partial \cdot C_\nu + 2n C^\nu \partial \cdot \phi_\nu \\ \nonumber
&&- 2 n (n-1) F \partial \cdot B + 2n C^\nu \partial_\nu A + 2 E \partial^\nu \phi_\nu
- 2n (n-1) F \partial^\nu D_\nu
 \\ \nonumber
&&+nC^\nu C_\nu + E^2 - n (n-1)F^2,
\eea
where $\Box=\partial_\mu\partial^\mu$, $\partial \cdot$ stands for the divergence with respect to the index $\mu_i$ and the contractions of the indices  $\mu_i$ are implicit.

From the Lagrangian \p{tn} one obtains the following equations of motion
\bea\label{eomB1}
&& \Box \phi_\nu + \partial C_\nu + \partial_\nu E = 0, \\ \label{eomB2}
&&\Box D_\nu + \partial \cdot C_\nu + \partial_\nu F =0, \\ \label{eomB3}
&&\Box A - \partial \cdot E + \partial F = 0, \\ \label{eomB4}
&& \Box B - \partial^\nu C_\nu =0, \\ \label{eomB5}
&& C_\nu - \partial D_\nu + \partial \cdot \phi_\nu  + \partial_\nu A=0, \\ \label{eomB6}
&& E + \partial B + \partial^\nu \phi_\nu = 0, \\ \label{eomB7}
&& F + \partial \cdot B +\partial^\nu D_\nu =0.
\eea

Finally, the gauge transformations \p{NSNSGT} take the following form
\bea\label{bgs1} 
&& \delta \phi_\nu = \partial \lambda_\nu + \partial_\nu \rho^{(1)}, \\  \label{bgs2}
&& \delta D_\nu = \partial \cdot \lambda_\nu + \partial_\nu \rho^{(2)}, \\  \label{bgs3}
&& \delta A = \partial  \rho^{(2)} - \partial \cdot \rho^{(1)}  - \tau, \\ \label{bgs4}
&& \delta B = - \partial_\nu \lambda^\nu + \tau, \\ \label{bgs5}
&&  \delta C_\nu  = - \Box \lambda_\nu + \partial_\nu \tau ,\\ \label{bgs6}
&& \delta E = - \Box \rho^{(1)} - \partial \tau, \\ \label{bgs7} 
&& \delta F = - \Box \rho^{(2)} - \partial \cdot \tau.
\eea

\section{Supersymmetry} \setcounter{equation}0 \label{Supersymmetry}
Before discussing the form of the supersymmetry transformations, let us convince ourselves that in the system under consideration the number of the bosonic and fermionic physical on-shell degrees of freedom match. The detailed proof is given in Appendix B, while below we only sketch the procedure. 

Let us split the vectorial index $\mu$ into transversal $i=1,..,D-1$ and longitudinal light-cone $(+,-)$ components.

In the bosonic sector, as one can see from the gauge transformation rules
\p{bgs1}-\p{bgs7}, one can use the gauge transformation parameters $\lambda_\nu$ and $\rho^{(1)}$ to 
gauge away all $"+"$ components in the field $\phi_\nu$. 
 The parameter
$\rho^{(2)}$ can be used to 
gauge away all the $"+"$ components in the field $D_\mu$ and the parameter $\tau$  to gauge away the field $B$. 
Then all the other fields $A$, $C_\mu$, $E$, $F$, $D_i$ and $D_-$  are equal to zero because of the field
equations \p{eomB1}-\p{eomB7}. 
This results in the transversality conditions for the field $\phi_\nu$
\be\label{TC-1}
\partial^{\mu_1} \phi_{\nu,\mu_1\mu_2\ldots\mu_n}=0,\qquad \partial^{\nu} \phi_{\nu,\mu_1\ldots\mu_n}=0.
\ee
which eliminate the $"-"$ components of $\phi$. Therefore,  one is  left
with the field which contains only
physical polarizations and obeys the Klein-Gordon equation
\be\label{KG-1}
\Box \phi_{j,i_1\ldots i_n}=0,
\ee
The index $j$ takes $D-2$ values and the symmetric tensor part of $\phi$ labelled by the indices $i$ has 
\be \label{d}
d=\frac {(D-3+n)!}{n! (D-3)!}
\ee
independent components.
 Therefore, the total number
of the physical bosonic degrees  freedom is $(D-2)d$.

The fermionic sector can be considered in a similar way. As one can see from the gauge transformation rule
\p{gtr-f-1},
using the parameter $\tilde \lambda$
one can gauge away the $"+"$ components of the physical field $\Psi$. Then the fields
$\Sigma$ and $\chi$ are zero because of the equations of motion \p{fermitriplet}. As a result
the physical field obeys the transversality condition
\be\label{TC-2}
\partial^{\mu_1} \Psi_{\mu_1\mu_2\ldots\mu_n}=0,
\ee
which eliminates the $"-"$ components. Therefore, one is left with only physical polarizations
which satisfy the Dirac equation
\be\label{D-1}
\gamma^\mu \partial_\mu  \Psi_{i_1\ldots i_n}=0.
\ee
Now let us count the number of the physical degrees of freedom. The number of degrees
of freedom labelled by the indices $i$ is the same as in the  case
of the bosonic field $\phi$ and equals to \p{d}. Further,
the Majorana spinor in $D=3$ has one on-shell degree of freedom, the Majorana spinor in $D=4$
has two on--shell degrees of freedom, the symplectic Majorana-Weyl (or just Weyl) spinor in $D=6$ has
four on-shell degrees of freedom and the Majorana-Weyl spinor in $D=10$ has eight on-shell degrees of freedom.
These numbers are equal to the $D-2$ on-shell degrees of freedom associated  with the index $j$ of the bosonic field $\phi$, eq. \eqref{KG-1}.
Therefore, in the system under consideration, the number of the physical bosonic and fermionic degrees
of freedom match in $D=3,4,6$ and 10. Let us note that these space-time dimensions are also singled out (due to twistor-like relations) in the so-called tensorial (or hyper) space formulation of infinite reducible conformally-invariant higher-spin systems (see e.g. \cite{Bandos:2005mb} and \cite{Bandos:2005rr,Sorokin:2017irs} for a review).

Now, let us proceed with the identification of the supersymmetry transformations of the fields.
The simplest way to find the operator which satisfies \p{DW} is the following \cite{Kazama:1986cy}. Let us take an ansatz
\be \label{UG}
\mathcal Q=\langle 0_{NS}| \exp({A_{IJ} \Psi^I_{NS} \Psi^{J+}_R + B_{IJ} \Psi^I_{NS} \Psi^{J}_{NS} +  
C_{IJ} \Psi^{I+}_{R} \Psi^{J+}_{R}}) | \tilde 0_R \rangle\,
\ee
where $\Psi^I_{NS}$ stand for the annihilation operators in the NS sector 
except for $(\alpha_\mu^i, b^i, c^i)$, since the latter set is common for both sectors.
The oscillators $\Psi^{J+}_R $
are the creation operators in the R sector, again except for $(\alpha_\mu^{i+}, b^{i+}, c^{i+})$.
One more obvious requirement is that each
separate term in the expression \p{UG} should have the ghost number equal to zero.
The constants $A_{IJ}, B_{IJ} $ and $C_{IJ}$  in \p{UG} are then determined from the equation
\p{DW}.

In (\ref{UG}) the vacuum   $ | \tilde 0_R \rangle$   in the R sector  
differs from the vacuum defined by \p{vacuumr-1}--\p{vacuumr-3}.
In particular, instead of \p{vacuumr-3}
we have  $| \tilde 0 \rangle_R$ having the following properties 
\be \label{vacuumr-4}
\gamma_0 | \tilde 0 \rangle_R=0
\ee
\be
(| \tilde 0 \rangle_R )^+ = \langle 0 _R |, \quad (|  0 \rangle_R )^+ = \langle  \tilde 0 _R |
\ee
\be
\langle 0 _R || 0_R \rangle = \langle \tilde 0 _R || \tilde 0_R \rangle =1.
\ee
The presence of the two Clifford vacua in the R sector is consistent and moreover is required for obtaining  the generator of the supersymmetry transformations (see \cite{Kazama:1986cy} for a detailed discussion). As we shall see later, the final answer for supersymmetry transformations
will be given
in terms of the ``proper" R vacuum defined in \p{vacuumr-1}--\p{vacuumr-2}, with the bosonic ghost zero modes
$\beta_0$ and $\gamma_0$ being completely eliminated. 

Now let us recall that for the truncated systems under consideration, in the R sector the fields  are created only by
$\alpha_1^{\mu +}, b_1^+$ and  $c_1^+$, whereas in the NS sector the fields are also created
by $\psi_\frac{1}{2}^{\nu +}, \beta_\frac{1}{2}^+$  and $\gamma_{\frac{1}{2}}^+$.

For such truncated systems the BRST charges in the R and NS sector  are
  \be \label{BRSTR-1} 
Q_{R} = \gamma_{0} g_{0} -\frac{1}{2}\gamma_0^2 \beta_0 +
Q^\prime,
\ee
 \be \label{BRSTNS-1} 
Q_{NS} = \gamma_{1/2} g^+_{1/2} + \gamma^+_{1/2} g_{1/2} - \gamma^+_{1/2} \gamma_{1/2} b_0 +
Q^\prime,
 \ee 
 where
 \be
 Q^\prime = c_0 l_0+  c_1^+ l_1 + c_1 l_1^+ - c_1^+ c_1 b_0.
 \ee
The operator $\mathcal Q$  takes the form
\be \label{UAN}
\mathcal Q=\langle 0_{NS}| \exp( \psi_0 \cdot \psi_{1/2} + \frac{i}{2} \gamma_{1/2} \beta_{1/2} - i \gamma_{1/2} \beta_0) | \tilde 0_R \rangle 
\ee
Obviously, the operator $Q^\prime$ commutes with $\mathcal Q$.
Then 
it is straightforward to check that \p{UAN}
indeed satisfies \p{DW}, provided that the constant coefficients in each term in the exponential are
chosen as in \p{UAN}.

Now using the explicit form \p{UAN} one can find the supersymmetry transformations
of the fields of the system.
To this end,
let us expand the operator \p{UAN} in the power series of $\beta_0$
\be
\mathcal Q(\beta_0) = \mathcal Q(0) + \beta_0 \mathcal Q^\prime(0) + ...
\ee
and use the equation \p{LL} as well as explicit forms of the higher spin functionals
in the NS \p{HSFNS} and R sector \p{truncated}. Since  we are dealing with the truncated HS functional, in the R sector we only need the terms given above, so one gets 
\begin{equation}\label{STR1}
\delta |\Phi^{NS}_1 \rangle = u^+ \, \epsilon^+ |\Phi^R_1 \rangle -
\gamma_{1/2}^+ u^+ \, \epsilon^+ |\Phi_2^R \rangle, \qquad \delta |\Phi^{NS}_2
\rangle = 2 u^+ g_0 \, \epsilon^+|\Phi^R_2 \rangle,
\end{equation}
\begin{equation}\label{STR2}
\delta |\Phi^{R}_1 \rangle = -2 \epsilon \,\,  g_0 u |\Phi^{NS}_1 \rangle +
\epsilon \,  \gamma_{1/2} u |\Phi_2^{NS} \rangle, \qquad \delta |\Phi^{R}_2
\rangle = \epsilon \, u  |\Phi^{NS}_2 \rangle
\end{equation}
with
\begin{equation}\label{w}
u = \langle 0^{NS}|\ \exp (\psi^\mu_0 \psi_{\mu 1/2} +
\frac{i}{2}\gamma_{1/2} \beta_{1/2} ) |0^R\rangle.
\end{equation}
Then one can show that the total Lagrangian
\begin{eqnarray}   \label{LT} \nonumber
{\cal L}_{tot.}&=&  \langle \Phi^{NS}_1|l_0|\Phi^{NS}_1\rangle -
 \langle \Phi^{NS}_2|\tilde Q_{NS}|\Phi^{NS}_1\rangle -
 \langle \Phi^{NS}_1|\tilde Q_{NS}|\Phi^{NS}_2\rangle +
 \langle \Phi^{NS}_2|M_{NS}|\Phi^{NS}_2\rangle \\
&&+
 \langle \Phi^{R}_1|g_0|\Phi^{R}_1\rangle +
 \langle \Phi^{R}_2|\tilde Q_{R}|\Phi^{R}_1\rangle +
 \langle \Phi^{R}_1|\tilde Q_{R}|\Phi^{R}_2\rangle -
2 \langle \Phi^{R}_2|M_R g_0|\Phi^{R}_2\rangle
\end{eqnarray}
is invariant under the supersymmetry transformations \p{STR1} and \p{STR2}.
Having found the form of the supersymmetry transformations of the higher-spin functionals, one can extract
from them the transformations of the component fields. To this end, one should expand the equations \p{STR1}--\p{STR2}
in the power series of the remaining ghost variables, as will be considered in the next Section for the cases of supermultiplets of different maximal spin.

\section{Supermultiplets} \setcounter{equation}0 \label{Supermultilets}
 
 \subsection{ Vector multiplet }
 The simplest example is the Maxwell supermultiplet for which there are no states associated with the creation operators $\alpha^{+\mu}_1$ (i.e. $n=0$).
 
As one can see from \p{NStrip1}, the fields in the bosonic NS sector are
\be
| \phi \rangle = A_\nu(x) \psi^{\nu +}_\frac{1}{2} |0  \rangle_{NS}
\ee
and
\be
| E\rangle = \beta_{\frac{1}{2}}^+E(x) |0  \rangle_{NS}.
\ee
Similarly, from  \p{Rtrip}  in the fermionic R sector  we have
\be
| \Psi \rangle = \Psi(x) |0 \rangle_{R}.
\ee
The Lagrangian   of the NS and R  sector is
\be \label{1-1}
L=-  A^\mu \Box A_\mu + E \partial^\mu A_\mu - A^\mu \partial_\mu E + E^2
 -i{\bar \Psi} \gamma^\mu \partial_\mu \Psi.
\ee
The parameter of the gauge transformations is
\be
| \rho \rangle= \rho(x)| 0 \rangle_{NS}
\ee
and the gauge transformations are
\be
\delta  A_\mu = \partial_\mu \rho, \qquad \delta E = - \Box \rho.
\ee
Expressing  the field $ E $ via its  equation of motion
\be \label{3}
E=- \partial^\mu A_\mu
\ee
and putting it back to \p{1-1} one obtains a conventional Maxwell Lagrangian
\be \label{4}
L_{NS}=-  A^\nu \Box A_\nu + A^\mu \partial_\mu \partial_\nu A^\nu=\frac 12 F^2_{\mu\nu}+total\,\,derivative\,\,.
\ee
The Lagrangian \p{1-1} is invariant under the following supersymmetry transformations of the fields. 
The supersymmetry transformation of the gauge field is
\be \label{ph}
\delta A_\mu = {i}\bar \Psi\gamma_\mu\epsilon
\ee
and that of $E$ is
\be
\delta E = - \partial^\mu \delta A_\mu   + {i}\bar \Psi\gamma^\mu \partial_\mu \epsilon=0
\ee
which is in accordance with what one obtains from the equation \p{STR1}.
For the fermionic field, from \p{STR2} we get
\be \label{5}
\delta  \Psi= -  \epsilon \gamma^\nu  \gamma^\mu \partial_\nu A_\mu - \epsilon E\,.
\ee
Let us note that one can  rewrite the Lagrangian \p{1-1} in the following form  (modulo total derivatives)
\be\label{1.1}
L=-  A^\mu \Box A_\mu + A^\mu \partial_\mu \partial_\nu A^\nu+ (E +\partial^\mu A_\mu)^2 -i{\bar \Psi} \gamma^\mu \partial_\mu \Psi.
\ee
Now one can introduce  the field
\be
{\cal D}=E+\partial^\mu A_\mu
\ee
which transforms as a total derivative under the supersymmetry transformations
\be
\delta {\cal D}= {i}\partial^\mu \bar \Psi_a (\gamma_\mu)^a{}_b \epsilon_b
\ee
and is the conventional
auxiliary ${\cal D}$-field of the  $N=1$, $D=4$  Maxwell  supermultiplet. 

As is well known, the algebra of the supersymmetry transformations \eqref{ph} and \eqref{5}, with $E$ satisfying \eqref{3}, closes (on the mass shell) on the translations and gauge transformations of the fields in space-time dimensions $D=3,4,6$ and 10.

\subsection{Gravitational and antisymmetric tensor supermultiplet}\label{gr+b}

The supersymmetric system of fields with the highest spin $s=2$, contained in \eqref{Rtrip} and \eqref{NStrip}, is generated by one creation operator $\alpha^{+\mu}_1$, hence $n=1$.

From the equation \p{NStrip} one concludes that the bosonic NS sector contains the following fields. The field
$\phi_{\nu,\mu}$ with no symmetry between the indices $\mu$ and $\nu$, 
two scalars $A(x)$ and $B(x)$, one of which is a pure gauge, and two vector auxiliary fields
$C_\nu(x)$ and $E_\mu(x)$. The fermionic Ramond sector contains two fields $\Psi_\mu(x)$ and $\chi(x)$.

The Lagrangian of the bosonic sector has the form
\bea \label{t2}
L_{NS}&&=  - \phi^{\nu, \mu} \Box \phi_{\nu,\mu } + B\Box A  + A \Box  B \\ \nonumber
&& +E^\mu  \partial_\mu B  + C^{\nu}\partial^\mu \phi_{\nu,\mu } + C^\nu \partial_\nu  A
+ E^\mu\partial^\nu
\phi_{ \nu, \mu } \\ \nonumber
&&- B \partial_\alpha E^\mu - \phi^{\nu, \mu } \partial_\mu C_\nu  -A \partial_\nu C^\nu - \phi^{\nu \mu}
\partial_\nu E_\mu \\ \nonumber
&&+C^\nu C_\nu + E^\mu E_\mu\,.
\eea
The equations of motion of $C_\nu$ and   $E_\mu$  are algebraic and allow one to express these fields as derivatives of $\phi_{ \nu, \mu }$, $A$ and $B$
\be \label{cc}
\hat C_\nu\equiv C_\nu + \partial^\mu\phi_{ \nu, \mu } + \partial_\nu A =0,
\ee
\be \label{ee}
\hat E_\mu \equiv E_\mu + \partial_\mu B + \partial^\nu \phi_{\nu, \mu }=0\,.
\ee
The Lagrangian \p{t2} is invariant under the gauge transformations with three parameters
$\lambda_\nu$, $\rho_\mu^{(1)}$ and $\tau$
\bea \nonumber
&&\delta \phi_{\nu, \mu } = \partial_\mu \lambda_\nu + \partial_\nu \rho_\mu^{(1)}, \\ \nonumber
&&\delta A = - \partial^\nu \rho_\nu^{(1)} - \tau, \\ \nonumber
&&\delta B= - \partial^\nu \lambda_\nu+ \tau, \\
&&\delta C_\nu  =- \Box  \lambda_\nu + \partial_\nu \tau, \\ \nonumber
&&\delta E_\mu=- \Box  \rho_\mu^{(1)}-\partial_\mu \tau.
\eea
The fermionic sector is described by the following Lagrangian
\be \label{Rs2-a}
L_R =-i  \bar \Psi^\mu \gamma^\nu \partial_\nu \Psi _\mu -i \bar \Psi^\mu \partial_\mu\chi +
i\bar \chi \partial_\mu \Psi^\mu+  i \bar \chi \gamma^\nu \partial_\nu \chi,
\ee
which is invariant under the gauge transformations
\be
\delta \Psi_\mu^a = \partial_\mu \tilde \lambda^a,  \qquad 
\delta \chi^a = -(\gamma^\mu)^a{}_b \partial_\mu \tilde \lambda^b.
\ee

The Lagrangian
\be
L=L_{MS}+L_R,
\ee
which is the sum of \p{t2} and \p{Rs2-a}, is invariant under the supersymmetry transformations
\bea \label{GR-TR-1}
&&\delta  \phi_{ \nu, \mu } = i \bar \Psi_{\mu} \gamma_\nu \, \epsilon, \\ \nonumber
&& \delta  C_{ \nu}=  -i (\partial_\mu \bar \chi)  \gamma^\mu \gamma_\nu\, \epsilon, \\ \nonumber
&& \delta  B = -i\bar \chi \, \epsilon, 
\eea
\bea \label{GR-TR-2}
&&\delta \Psi_\mu = - \gamma^\nu\gamma^\rho\epsilon \,\partial_\nu \phi_{\rho,\mu}
-  \epsilon E_\mu,
 \\ \nonumber
&&\delta \chi= - \gamma^\nu\epsilon\,  C_\nu\,. 
\eea
With the use of the equation of motion \p{ee} for the field $E_\mu$, the supersymmetry transformation rule
for $\delta \Psi_\mu$ can be equivalently rewritten as
\be
\delta \Psi_\mu=-2 \gamma^{\nu \rho}\epsilon \,\partial_\nu \phi_{\rho, \mu } + \epsilon
\partial_\mu B,
\ee
where $\gamma^{\nu\rho}=\gamma^{[\nu}\gamma^{\rho]}$.
Using the Fierz identities \p{GG-1}-\p{FI}, one can check that the algebra of supersymmetry transformations closes on the translations of the fields modulo their field-dependent gauge transformations (and equations of motion). 

As one can see from the equations above, the $N=1$ supermultiplet under consideration is reducible or irreducible depending on the dimension of space-time. To see this, let us rewrite the above Lagrangian in terms of completely decoupled fields transforming under irreducible representations of the Lorentz group.

\subsubsection{Diagonalization of the Lagrangians}

To diagonalize the Lagrangians \eqref{t2} and \eqref{Rs2-a} and make them be the sum of the free Lagrangians for the Lorentz-group irreducible fields we introduce the gauge invariant scalar field  
\be\label{varphi}
\varphi=A+B+\phi',
\ee
where $\phi'=\eta^{\nu\mu}\phi_{\nu,\mu}$, and the spinor field
\be\label{hatchi}
\hat\chi=\chi+\gamma^\mu\Psi_\mu\,.
\ee
Then we redefine the field $\phi_{\nu,\mu}$ as follows
\be\label{phinew}
\phi_{\nu,\mu}=g_{\nu\mu}+B_{\nu\mu}-\frac 12\eta_{\nu\mu}(A+B+g')=g_{\nu\mu}+B_{\nu\mu}-\frac 1{2-D} \eta_{\nu\mu}\varphi,
\ee
where $g_{\nu\mu}$ is symmetric and $B_{\nu\mu}$ is antisymmetric.

The field $\Psi_{\mu}$ is redefined as follows
\be\label{Psinew}
\Psi_{\mu}=\hat \Psi_\mu-\frac 12\gamma_\mu(\chi+\gamma^\nu\hat \Psi_{\nu})=\hat \Psi_\mu-\frac 1{2-D}\gamma_\mu \hat\chi.
\ee
Note also that 
\be\label{chi}
\chi=\frac 2{2-D}\hat\chi-\gamma_\mu\Psi^\mu.
\ee
For the newly introduced fields and with the expressions \eqref{cc} and \eqref{ee}  for the auxiliary fields taken into account, the supersymmetry transformations take the following form (modulo gauge transformations)
\bea\label{susyB}
&&\delta  g_{  \nu \mu } = \bar {\hat\Psi}_{(\nu} \gamma_{\mu)} \,\, \epsilon, \\ \nonumber
&&\delta  B_{  \nu \mu } = \bar {\hat\Psi}_{[\nu} \gamma_{\mu]} \,\, \epsilon-\frac 12 \bar{\hat\chi}\gamma_{\nu\mu}\epsilon,\\ \nonumber
&& \delta  \varphi = -\bar \epsilon  {\hat \chi} ,
\eea
\bea\label{susyPsi}
&&\delta \hat \Psi_\mu = i \gamma^{\nu\rho}\,\epsilon \, \partial_\nu g_{\mu \rho}-\frac i2\gamma^{\nu\rho}\,\epsilon \, H_{\mu\nu\rho}+\frac i{3(D-2)}\gamma_\mu\gamma^{\lambda\nu\rho}H_{\lambda\nu\rho},
 \\ \nonumber
&& = i \gamma^{\nu\rho}\,\epsilon \,\, \partial_\nu g_{\mu\rho}+\frac i{3(D-2)}\left(\gamma_\nu{}^{\lambda\nu\rho}-\frac{3(D-4)}{2}\delta_\mu^\rho\gamma^{\lambda\nu}\right)H_{\lambda\nu\rho},
 \\ \nonumber
&&\delta \hat\chi= -i \gamma^\mu \, \epsilon\,  \partial_\mu\varphi-\frac i3 \gamma^{\mu\nu\rho}\epsilon H_{\mu\nu\rho},
\eea
where $H_{\mu\nu\rho}=3\partial_{[\mu}B_{\nu\rho]}$.

The factorized Lagrangians for the newly defined fields have the following form
\bea\label{BL}
L_{NS}&&= \frac 13 H_{\lambda\mu\nu}H^{\lambda\mu\nu} - g^{\nu \mu} \Box g_{\nu \mu} + 2g^{\nu\mu}\partial_\nu \partial^\rho g_{\rho\mu}-2g^{\nu\mu}\partial_\nu\partial_\mu g'+g'\Box g'
\\ \nonumber
 &&-\frac 1{(D-2)} \varphi\Box\varphi 
+\hat C^\mu \hat C_\mu + \hat E^\nu \hat E_\nu,
\eea
(where $C_\mu$ and $E_\nu$ were defined in \eqref{cc} and \eqref{ee}) and 
\bea\label{FL}
L_R &=& -i  \bar{\hat\Psi}^\nu \gamma^\mu \partial_\mu \hat\Psi _\nu +i\bar{\hat\Psi}^\nu  \partial_\nu (\gamma^\mu\hat\Psi _\mu)+i\bar{\hat\Psi}^\mu\gamma_\mu  \partial_\nu \hat\Psi^\nu-i(\bar{\hat\Psi}\gamma) \slash\!\!\!\partial(\gamma{\hat\Psi})-\frac i{D-2} \bar{\hat \chi} \gamma^\mu \partial_\mu \hat\chi\nonumber\\
&=&-i\bar{\hat\Psi}_\mu\gamma^{\mu\nu\lambda}\partial_\nu{\hat\Psi}_\lambda-\frac i{D-2} \bar{\hat \chi} \gamma^\mu \partial_\mu\hat\chi.
\eea
The Lagrangian for the symmetric field $g_{\mu\nu}$ is nothing but the Fierz-Pauli Lagrangian for the massless spin-2 field, while the Lagrangian for the fermionic field $\hat\Psi_\mu$ is the Rarita-Schwinger Lagrangian for the massless spin-3/2 field.

\subsubsection{$D=4$}
In $D=4$ the fields $g_{\mu\nu }$ and $\hat\Psi_\mu$ form the linearized $N=1$ supergravity multiplet.
The anti-symmetric field $B_{ \mu\nu}$, the scalar field $\varphi$ and the spinor $\hat\chi$ form the $N=1$ linear supermultiplet. Here let us note a somewhat unconventional form of the supersymmetry transformations \eqref{susyB} and \eqref{susyPsi} which naively seem to transform into each other the fields $B_{\mu\nu}$ and $\hat\Psi_\mu$ which belong to different $N=1$, $D=4$ supermultiplets. However, as one can show, these terms in the supersymmetry transformations of  $B_{\mu\nu}$ and $\hat\Psi_\mu$ are trivial in the sense that their commutator is identically zero (in $D=4$). Hence, in $D=4$ these terms are redundant and can be skipped \footnote{We are grateful to Sergei Kuzenko for clarifying discussion of this issue.}.

\subsubsection{$D=6$}
In $D=6$ the reducible system under consideration is composed of the $N=(1,0)$ gravitational multiplet ($g_{\mu\nu}$, $B^+_{\mu\nu}, \hat \Psi_\mu$), where $B^+_{\mu\nu}$ is the self-dual part of $B_{\mu\nu}$ (i.e. $H^+_3=dB^+_2=*H_3^+)$ and $\hat \Psi_\mu$ is a chiral spinor, and the $N=(1,0)$ tensor supermultiplet ($B^-_{\mu\nu},\hat \chi$) in which $B^-_{\mu\nu}$ is the anti-self-dual part of $B_{\mu\nu}$ and $\hat\chi$ is an anti-chiral spinor.

\subsubsection{$D=10$}
In this case, the  bosonic fields $g_{\mu\nu}$, $B_{\mu\nu}$, 
$\varphi$  and the Majorana-Weyl spinor fields $\hat \Psi_\mu$ and 
$\hat \chi$  of the opposite chirality form
the irreducible supermultiplet of the linearized $N=1$, $D=10$ supergravity.

\subsection{Higher-spin Supermultiplets}
As we discussed in detail in Sections \ref{Fieldcontent} and 4, in the generic case the supersymmetric higher-spin system is formed by the generalized bosonic `triplet' consisting of the mixed-symmetry fields $\phi_{\nu,\mu_1\ldots\mu_n}$, $C_{\nu,  \mu_1... \mu_{n-1}}$ and $D_{\nu,  \mu_1... \mu_{n-2}}$, which are symmetric tensors in the indices $\mu_i$ but with no symmetry between $\nu$ and $\mu_i$, and the symmetric tensor fields $A_{\mu_1...\mu_{n-1}}$, $B_{\mu_1... \mu_{n-1}}$ and $F_{\mu_1... \mu_{n-2}}$, and by the fermionic triplet consisting of the symmetric spinor-tensor fields $\Psi_{\mu_1...\mu_{n}}$, $\chi_{\mu_1...\mu_{n-1}}$ and $\Sigma_{\mu_1...\mu_{n-2}}$. The bosonic fields $\phi, C$ and $D$ can be decomposed into the irreducible representations of the $GL(D)$ group (which are still reducible with respect to $SO(1,D-1)$ since they are traceful). In particular, one obtains
fields described by totally symmetrical Young diagrams and the fields  with mixed symmetries (``the hook"). The latter
are described by the Young diagrams with two rows, with only one box in the second row.

The Lagrangian describing this higher-spin system is the sum of the Lagrangians \eqref{RR-2} and \eqref{tn}. It is invariant under the following supersymmetry transformations of the fields

\bea \label{MAIN1}
&&\delta \Psi_{\mu_1... \mu_n} 
=-  \gamma^\rho\gamma^\nu\,  \epsilon \, \partial_\rho 
\phi_{\nu, \mu_1...\mu_n }
- \epsilon E_{\mu_1... \mu_n},
 \\ \nonumber
&&\delta \Sigma_{\mu_1,..., \mu_{n-2}} = - \gamma^\rho\gamma^\nu
\, \epsilon \, \partial_\rho  D_{\nu, \mu_1... \mu_{n-2} }
- \epsilon F_{\mu_1... \mu_{n-2}},
 \\ \nonumber
&&\delta \chi_{\mu_1... \mu_{n-1}}= - \gamma^\nu\, \epsilon\,C_{ \nu, \mu_1,... \mu_{n-1} },
\eea
\bea \label{MAIN2}
&&\delta  \phi_{\nu,  \mu_1... \mu_n } =i \bar \Psi_{\mu_1,..., \mu_n}
\gamma_\nu\, \epsilon, \\ \nonumber
&&\delta  D_{ \nu, \mu_1... \mu_{n-2} }
= i\bar \Sigma_{\mu_1... \mu_{n-2}}\gamma_\nu\, \epsilon, \\ \nonumber
&& \delta  C_{\nu,  \mu_1... \mu_{n-1} }= -  i \partial_\rho \bar 
\chi_{\mu_1,..., \mu_{n-1}}  \gamma^\rho\gamma_\nu\, \epsilon,\\ \nonumber
&& \delta  B_{\mu_1... \mu_{n-1}} = -i\bar \chi_{ \mu_1...\mu_{n-1}} \, \epsilon, \\ \nonumber
&& \delta  A_{\mu_1... \mu_{n-1}} = \delta  E_{\mu_1... \mu_{n}} = \delta  F_{\mu_1... \mu_{n-2}}=0.
\eea
Again, using the Fierz identities \p{GG-1}--\p{FI}, one can check that in $D=3,4,6$ and 10 the algebra of the supersymmetry transformations closes on the translations of the fields modulo their field-dependent gauge transformations and equations of motion. The easiest way to verify the on-shell closure of the supersymmetry algebra is to impose on the fields the light-cone gauge discussed in detail in Appendix B. In this gauge the supersymmetry transformations of the physical modes reduce to
\be\label{gft}
\delta \Psi_{i_1... i_n} 
=-  \gamma^{\rho}\gamma^{\nu}\,  \epsilon \, \partial_\rho 
\phi_{\nu, i_1\ldots i_n },\qquad
\delta  \phi_{\nu,  i_1... i_n } =i \bar \Psi_{i_1,..., i_n}
\gamma_\nu\, \epsilon.
\ee
Note that the form of these transformations is similar to that of the vector multiplet, eq. \eqref{ph} and \eqref{5} (in the Lorentz gauge), with the indices $i_r$ being purely external.

Finally, the physical fields $\phi_{j,i_1\ldots i_n}$ can be decomposed into the symmetric field $\varphi_{i_1\ldots i_{n+1}}$ and the mixed symmetry field $B_{j,i_1\ldots i_n}$ such that $B_{(j,i_1\ldots i_n)}$=0. The field $\varphi_{i_1\ldots i_{n+1}}$ contains the irreducible traceless symmetric fields of the ranks (spins) $s=n+1$, $n-1$, $n-3$ \ldots, 1 (or 0). The field $B_{j,i_1\ldots i_n}$  further decomposes into the traceless mixed symmetry fields  $b_{j,i_1\ldots i_r}$ with $r=n,n-2,\ldots, 1$ (or 0).

The fermionic field $\Psi_{i_1,..., i_n}$ decomposes into the irreducible gamma-traceless fields $\psi_{i_1,..., i_r}$ with $r=n,n-1,\ldots,0$ and corresponding spins $s=n+\frac 12,n-\frac 12,\ldots,\frac 12$.

In $D=4$ the mixed symmetry fields $B_{j,i_1\ldots i_n}$ can be dualized into totally symmetric fields $\tilde \varphi_{i_1\ldots i_{n-1}}$ which together with $\varphi_{i_1\ldots i_{n+1}}$ and $\Psi_{i_1,..., i_n}$ form reducible on-shell $N=1$ supermultiplets that  split into the irreducible supersymmetric sets with spins $(s,s-\frac 12)$ and $(s-2,s-\frac  32)$ with $s=1,\ldots, n+1$.  Irreducible D=4 higher-spin supermultiplets first appeared in \cite{Curtright:1979uz} and generalized to off-shell supermultiplets in \cite{Kuzenko:1993jp,Kuzenko:1993jq} (see also \cite{Buchbinder:1995uq} for a review).

\section{Conclusion} \setcounter{equation}0 \label{Conclusions}
In this paper we have considered supersymmetry between reducible higher-spin multiplets in space-times of dimension $D=3,4,6$ and $10$.
It is straightforward to obtain corresponding supersymmetric systems of massive higher-spin multiplets by performing the
BRST dimensional reduction of the massless ones as in \cite{Pashnev:1997rm}.
Another possible continuation of this work is the generalization of the considered higher-spin systems to $D$-dimensional $AdS$ spaces using the results of \cite{Agugliaro:2016ngl} for the fermionic triplets
and generalizing results of \cite{Sagnotti:2003qa} to mixed symmetry fields on $AdS_D$. In $AdS_4$ such a generalization should reproduce in components some of the irreducible higher-spin superfield constructions of \cite{Kuzenko:1994dm}. 
A generalization to the systems with extended supersymmetries is yet another interesting issue, as well as understanding whether and how the supersymmetric higher-spin systems considered above fit into higher-spin superalgebras \cite{Fradkin:1989ywa,Fradkin:1989yd,Vasiliev:2004cm,Alkalaev:2010af,Sezgin:2012ag}.
Finally, it is important to find a deformation of the operator of supersymmtry transformations
\p{w} in such a way that it is consistent with interacting higher spin theories.
\footnote{ These theories extensively studied recently up to the quartic order in fields contain cubic and quartic vertices, 
(see
\cite{Metsaev:1991mt, Metsaev:1991nb, Fotopoulos:2010ay, Polyakov:2010sk, Taronna:2011kt, Dempster:2012vw,
McGady:2013sga,Ponomarev:2016jqk,Ponomarev:2016lrm,
Bengtsson:2016hss,Taronna:2017wbx,Roiban:2017iqg}), and the deformation of the supersymmetry operator \p{w} should be consitent with these vertices.}

\vskip 0.5cm

\noindent {\bf Acknowledgments.} We are
grateful to K. Alkalaev,
C. Angelantonj, I. Bandos, X. Bekaert, I.L. Buchbinder, A. Fotopoulos,  V.A. Krykhtin, S.M. Kuzenko,
A.Sagnotti and I. Samsonov 
for many illuminating  discussions.
D.S. is grateful to the Department of Physics, UWA and the School of Mathematics, the University of Melbourne for hospitality during the final stage of this work.
M.T. is grateful  to the Theory Division of CERN, Switzerland and to the
Department of Mathematics, the University of Auckland, New Zealand,  where various stages of this project  were carried out. 
This work
was supported by the Australian Research Council grant DP160103633.
Work of D.S. was also partially supported by the Russian Science
Foundation grant 14-42-00047 in association with Lebedev Physical Institute.

\renewcommand{\thesection}{A}

\renewcommand{\theequation}{A.\arabic{equation}}

\setcounter{equation}0
\appendix
\numberwithin{equation}{section}

\section{Conventions}\label{Appendix A}
Throughout the paper  ``$(,)$" denotes symmetrization and ``$[,]$" denotes antisymmetrization with weight one.

The $\gamma$--matrices satisfy the following anti-commutation relations
\begin{equation}\label{1}
(\gamma^\mu)^a{}_c (\gamma^\nu)^c{}_b
+
(\gamma^\nu)^a{}_c (\gamma^\mu)^c{}_b
 = 2 \eta^{\mu \nu} \delta^a_b \,,
\end{equation}
where $a$,  $b$ label spinorial indices,  and the Greek letters
$\mu, \nu, \ldots$ label space-time vector indices. The spinor indices can be raised and lowered by anti-symmetric charge conjugation matrices $C_{ab}$ and $C^{ab}$  $(C^{ab}C_{bc}=-\delta_a^c)$.

For checking the on-shell closure of the supersymmetry algebra we have used the following gamma-matrix identities
\be \label{GG-1}
(\gamma_{\mu \nu \rho})^a{}_b= (\gamma_{\mu \nu})^a{}_c (\gamma_\rho)^c{}_b +2 \eta_{\rho [\nu} (\gamma_{\mu]})^a{}_b,
\ee
\be\label{FI}
(\gamma^\nu)_{ab}{(\gamma_ \nu)}_{ cd}+ (\gamma^\nu)_{ac}{(\gamma_ \nu)}_{ db} + (\gamma^\nu)_{ad}{(\gamma_ \nu)}_{ bc}=0.
\ee

\section{Details of the light-cone gauge fixing procedure and the equality of the bosonic and fermionic degrees of freedom}

The gauge transformations of the reducible bosonic system are given in eqs. \eqref{bgs1}-\eqref{bgs7}. Let us split the space-time indices $\nu$ and $\mu_r$ ($r=1,\ldots, n$) into the $D-2$ transversal indices 
$j$ and $i_r$ and the light-cone indices $(+,-)$.

Consider now the gauge transformations \eqref{bgs1} of the field $\phi_{\nu,\mu_1\ldots\mu_n}$. These can be used to gauge away the components $\phi_{+,\mu_1\ldots\mu_n}$
\be\label{phi+,}
\phi_{+,\mu_1\ldots\mu_n}=0. 
\ee
Upon imposing this gauge the remaining components of $\phi_{\nu,\mu_1\ldots\mu_n}$, are transformed with the parameters $\lambda_{\nu, \mu_2\ldots m_n}$ and $\rho^{(1)}_{\mu_1\ldots \mu_n}$ related by
\be\label{r1}
\partial_+\rho^{(1)}_{\mu_1\ldots \mu_n}+n\partial_{(\mu_1}\lambda_{+,\mu_2\ldots \mu_n)}=0,
\ee
in which the symmetrization involves  the indices $\mu_r$ only. Therefore, the remaining components $\phi_{-,\mu_1\ldots\mu_n}$ and $\phi_{j,\mu_1\ldots\mu_n}$ are transformed by \eqref{bgs1} in which $\lambda_{\nu,\mu_2\ldots \mu_n}$ are independent parameters. These can be used to put to zero the following components
\be\label{phii+}
\phi_{-,+\mu_2\ldots\mu_n}=0,\qquad \phi_{i,+\mu_2\ldots\mu_n}=0
\ee
and impose the transversality condition
\be\label{tr}
\partial^\nu\phi_{\nu,\mu_1\ldots\mu_n}=\partial^j\phi_{j,-\mu_2\ldots \mu_n}+\partial_+\phi_{-,-\mu_2\ldots\mu_n}=0.
\ee
So, at this stage we are left with the following independent components of $\phi$
\be\label{phijhat}
\phi_{j,\hat i_1\ldots \hat i_n}\qquad (\hat i_r=(i_r,-))\,.
\ee
Now let us consider the transformations \eqref{bgs2} of the field $D_{\nu,\mu_1\ldots mu_{n-1}}$. The independent parameters $\rho^{(2)}_{\mu_1\ldots \mu_n}$ can be used to put to zero 
\be\label{D+}
\partial^\nu D_{\nu,\mu_1\ldots \mu_{n-1}}=0,
\ee
while the parameters $\tau_{\mu_1\ldots \mu_{n-1}}$ in \eqref{bgs2} can be used to gauge fix to zero the field $B$
\be\label{B0}
B_{\mu_1\ldots \mu_{n-1}}=0.
\ee
Let us now pass to the consideration of the equations of motion \eqref{eomB1}-\eqref{eomB7}. Upon the above gauge fixing they reduce to
\bea\label{eomB11}
&& \Box \phi_\nu + \partial C_\nu = 0, \\ \label{eomB21}
&&\Box D_\nu + \partial \cdot C_\nu =0, \\ \label{eomB31}
&&\Box A  = 0, \\ \label{eomB41}
&& C_\nu - \partial D_\nu + \partial \cdot \phi_\nu  + \partial_\nu A=0.  \label{eomB61}\\
&& E=F=0.
\eea
From these equations it also follows that 
\be\label{Ccc}
\Box C_\nu=0, \qquad \partial^\nu C_\nu=0\,.
\ee
Let us now consider eq. \eqref{eomB11}. From \eqref{phijhat} it follows that
\be\label{C0}
\partial_+C_{\nu,\mu_1\ldots \mu_{n-1}}+(n-1)\partial_{(\mu_1}C_{\mu_2\ldots \mu_{n-1})+,\nu}=0.
\ee
If in \eqref{C0} all the indices $\mu_r$ are $+$ we have
\be\label{C+}
\partial_+C_{\nu,+\ldots +}=0.
\ee
Since the field $C_\nu$ is on the mass shell \eqref{Ccc}, and assuming that its momentum $p_+$ (associated with $\partial_+$) is non-zero, eq. \eqref{C+} implies that
\be\label{C+0}
C_{\nu,+\ldots +}=0.
\ee
Then, if only $\mu_1$ is different from $+$, from \eqref{Ccc}, \eqref{C0} and \eqref{C+0} it follows that
\be\label{C+1}
\partial_+C_{\nu,\mu_1+\ldots +}=0 \quad \rightarrow \quad C_{\nu,\mu_1+\ldots +}=0.
\ee
Continuing this recursion procedure we finally find that 
\be\label{C+n}
C_{\nu,\mu_1\ldots \mu_{n-1}}=0,
\ee
and the equations of motion \eqref{eomB11}-\eqref{eomB61} further reduce to 
\bea\label{eomB12}
&& \Box \phi_\nu = 0, \\ \label{eomB22}
&&\Box D_\nu  =0, \\ \label{eomB32}
&&\Box A  = 0, \\ 
&&  - \partial D_\nu + \partial \cdot \phi_\nu  + \partial_\nu A=0.  \label{eomB62}
\eea
Since the field $D_\nu$ is on-shell \eqref{eomB32}, we can use the residual gauge transformation \eqref{bgs3} with the on-shell parameter $\rho^{(2)}$ (i.e. $\Box \rho^{(2)}=0$) to put to zero $D_+$
\be\label{D++}
D_{+,\mu_1\ldots\mu_{n-2}}=0.
\ee
From  \eqref{eomB32} and \eqref{eomB62} it then follows that
\be\label{A0}
\partial_+A=0 \quad \rightarrow \quad A=0
\ee
and hence
\be\label{D}
- (n-1)\partial_{(\mu_1} D_{\mu_2\ldots\mu_{n-1}),\nu} + \partial^{\mu_n}\phi_{\nu,\mu_1\ldots \mu_{(n-1)}\mu_n} =0.
\ee
Taking in \eqref{D} one of the indices $\mu$ be $+$ and remembering \eqref{phijhat}, we get
\be\label{D+++}
\partial_{+} D_{\mu_2\ldots\mu_{n-1},\nu}+(n-2)\partial_{(\mu_2} D_{\mu_3\ldots\mu_{n-1})+,\nu}=0,
\ee
which is similar to \eqref{C0}. Thus, its analysis results in
\be\label{D0}
D_{\nu,\mu_1\ldots \mu_{n-2}}=0.
\ee
Then eq. \eqref{D} reduces to the second transversality condition on $\phi$ whose non-zero components are give in \eqref{phijhat}
\be\label{tr2}
\partial^{\mu_n}\phi_{\nu,\mu_1\ldots \mu_{(n-1)}\mu_n} =0.
\ee
This condition further reduces the number of the independent components in $\phi$ which become those of the mixed symmetry tensor
\be\label{phii}
\phi_{j,i_1\ldots i_n}, \qquad j,i_r=1,\ldots, D-2.
\ee
Therefore, the bosonic ~reducible ~system ~under ~consideration ~has $N_B=(D-2)\frac{(D+n-3)!}{n!(D-3)!}$ physical degrees of freedom.

The gauge fixing procedure can be carried out in a slightly different way, similar to that  done for massless
\cite{Francia:2002pt} and massive \cite{Pashnev:1997rm} (non-generalized) bosonic triplets.
First one gauges away the fields $B$, $C_\mu$, $E$ and $F$ using 
the gauge parameters $\tau$, $\lambda_\mu$, $\rho^{(1)}$ and $\rho^{(2)}$, respectively.
Then one is left with the gauge freedom whose parameters satisfy the  equations
\be \label{restriction}
\Box \lambda_\mu=\partial_\mu \partial^\nu\lambda_\nu,\qquad  \Box \rho^{(1)}= -\partial(\partial^\nu\lambda_\nu),\qquad \Box \rho^{(2)}=-\partial \cdot (\partial^\nu\lambda_\nu).
\ee
The field equations reduce to
\bea\label{eomB1-A}
&& \Box \phi_\nu  = 0, \\ \label{eomB2-A}
&&\Box D_\nu  =0, \\ \label{eomB3-A}
&&\Box A  = 0, \\ 
\label{eomB5-A}
&&   \partial D_\nu - \partial \cdot \phi_\nu  - \partial_\nu A=0, \\ \label{eomB6-A}
&&  
\partial^\nu \phi_\nu = 0, \\ \label{eomB7-A}
&&  
\partial^\nu D_\nu =0. 
\eea
Now one  gauges away  the $"+"$
components of the field $\phi_\nu$ as well as the $D_+$ components, as has been done before.
Note that this gauge fixing is consistent with the restrictions \p{restriction} on the gauge parameters,
since the fields obey the Klein-Gordon equation \p{eomB1-A} and \p{eomB2-A}, and the transversality conditions \eqref{eomB6-A} and \eqref{eomB7-A}. 
Then from \eqref{eomB3-A} and \eqref{eomB5-A} it follows that $\partial_+A=0\,\rightarrow A=0$,
and \eqref{eomB5-A} reduces to \eqref{D} which implies $D_\nu=0$ and the second transversality condition \eqref{phijhat} on $\phi$.

The gauge fixing procedure in the fermionic sector is completely analogous. The result is that the independent physical components of the fermionic triplet can be associated with the spinor-tensor $\Psi_{a,i_1\ldots i_n}$ satisfying the Dirac equation. The number of independent components of this tensor in
$D=3,4,6$ and $10$ is the same as the number of the physical degrees of freedom of the bosonic sector, i.e. $N_F=(D-2)\frac{(D+n-3)!}{n!(D-3)!}$.


\providecommand{\href}[2]{#2}\begingroup\raggedright\endgroup

\end{document}